\documentclass[11pt]{article}
\usepackage[margin=1in]{geometry}
\usepackage{authblk}
\usepackage{times}
\usepackage[noadjust]{cite}
\usepackage[normalem]{ulem}
\usepackage[english]{babel}
\usepackage{blindtext}

\usepackage{booktabs}
\usepackage{url}
\usepackage{graphicx}
\usepackage{tikz}
\usepackage{amsmath}
\usepackage{epsfig,endnotes, graphicx,booktabs, multirow, array}
\usepackage{xspace}
\usepackage{colortbl}

\usepackage{tabularx}
\usepackage[utf8]{inputenc}
\usepackage{regexpatch}
\usepackage[font=footnotesize]{caption} %
\makeatletter
\newcommand{\linebreakand}{%
  \end{@IEEEauthorhalign}
  \hfill\mbox{}\par
  \mbox{}\hfill\begin{@IEEEauthorhalign}
}
\makeatother

\usepackage{algorithm}
\usepackage[noend]{algpseudocode}

\usepackage{subcaption}
\usepackage{xspace}
\usepackage{hyperref}
\usepackage[textsize=small,backgroundcolor=orange]{todonotes}
\usepackage{booktabs}
\usepackage{graphicx}
\usepackage{enumitem}
\usepackage{multirow,tabularx}

\usepackage{enumitem}
\usepackage{csquotes}
\usepackage{booktabs}
\usepackage{tabularx}
\usepackage{amsmath}
\usepackage{makecell}
\usepackage{multirow}
\RequirePackage[capitalise]{cleveref} %
\usepackage{cleveref}
\usepackage{caption}
\usepackage{subcaption}
\usepackage{epsfig,endnotes, graphicx,booktabs, multirow, array}

\definecolor{aliceblue}{rgb}{0.94, 0.97, 1.0}
\definecolor{maroon}{cmyk}{0,0.87,0.68,0.32}
\usepackage{tcolorbox}
\newtcolorbox{mybox}[1]{colback=aliceblue,colframe=black,fonttitle=\bfseries,title=#1}

\newcommand{\nnumber}[1]{\textcolor{black}{#1}}

\usepackage{tikz}
\usepackage{amsmath}
\usepackage{xspace}
\usepackage{colortbl}
\usepackage{authblk}
\usepackage{tabularx}
\usepackage[title]{appendix}
\usepackage[utf8]{inputenc}
\usepackage{mdframed}

\definecolor{aliceblue}{rgb}{0.94, 0.97, 1.0}
\definecolor{beaublue}{rgb}{0.74, 0.83, 0.9}

\definecolor{aliceblue}{rgb}{0.94, 0.97, 1.0}

\definecolor{aliceblue}{rgb}{0.94, 0.97, 1.0}

\usepackage{tcolorbox}

\hypersetup{
    colorlinks=true,
    linkcolor=blue,
    filecolor=magenta,      
    urlcolor=cyan
    }

\usepackage{cite}
\usepackage{breakurl}           %
\usepackage{url}                %
\usepackage{xcolor}             %
\usepackage[]{hyperref}         %
\hypersetup{                    %
  colorlinks,
  linkcolor={green!80!black},
  citecolor={red!70!black},
  urlcolor={blue!70!black}
}

\usepackage{mwe}
\usepackage{tikz}
\usetikzlibrary{arrows}
\usepackage{verbatim}

\usepackage{booktabs}
\usepackage{dirtree}
\usepackage{wrapfig}
\usepackage{caption}
\usepackage{subcaption}

\title{\Large \bf Unpacking Privacy Labels: A Measurement and Developer Perspective on Google's Data Safety Section}

\renewcommand*{\Affilfont}{\normalsize\normalfont}
\newsavebox\affbox

\author[1]{Rishabh Khandelwal\thanks{Equal Contribution}}
\author[1]{Asmit Nayak$^*$}
\author[1]{Paul Chung}
\author[1]{Kassem Fawaz}
\affil[1]{%
  \savebox\affbox{\Affilfont Department of Chemical Engineering, University of AAAAA BBBBBB, CCCCC road,}%
  \parbox[t]{\wd\affbox}{\protect\centering} University of Wisconsin -- Madison} 
\date{}

\begin{document}

\date{}

\maketitle
\begin{abstract}
     Google has mandated developers to use Data Safety Sections (DSS) to increase transparency in data collection and sharing practices. In this paper, we present a comprehensive analysis of Google's Data Safety Section (DSS) using both quantitative and qualitative methods. We conduct the first large-scale measurement study of DSS using apps from Android Play store (n=1.1M). We find that there are internal inconsistencies within the reported practices. We also find trends of both over and under-reporting practices in the DSSs. 
     Next, we conduct a longitudinal study of DSS to explore how the reported practices evolve over time, and find that the developers are still adjusting their practices. To contextualize these findings, we conduct a developer study, uncovering the process that app developers undergo when working with DSS. We highlight the challenges faced and strategies employed by developers for DSS submission, and the factors contributing to changes in the DSS. Our research contributes valuable insights into the complexities of implementing and maintaining privacy labels, underlining the need for better resources, tools, and guidelines to aid developers. This understanding is crucial as the accuracy and reliability of privacy labels directly impact their effectiveness.

\end{abstract}

\section{Introduction}
\label{sec:intro}

Privacy policies have traditionally served as the primary method for conveying the privacy practices of a service to users. However, studies have demonstrated that privacy policies are often ineffective, largely because users neglect them due to their length and vagueness ~\cite{cate2010limits, gluck2016short}. Introduced by Kelly et al. ~\cite{kelley_labels}, privacy nutrition labels summarize the privacy practices of websites in a nutrition label format, making them easier to understand. The concept of privacy labels has gained traction in the tech industry, with Google introducing Data Safety Sections (DSS) and Apple introducing Apple Privacy Labels (APL) for all new and updated apps on their app stores.

Recently, researchers showed the utility of privacy labels for users, making privacy practices more accessible~\cite{zhang2022usable}. However, the utility of the privacy labels depends on the developers properly filling out the forms. These forms should accurately reflect the developers' intentions regarding the privacy practices of their apps. Underreporting these practices can lead to inaccurate privacy labels, confusing and potentially instilling a false sense of security and privacy among users. Conversely, overreporting privacy practices can harm the developer by deterring users who will perceive the apps as less secure and private than they are.

Therefore, it is essential to investigate the developer's experience with reporting their practices via the privacy labels. Prior research has reported the responsiveness of developers in implementing Apple Privacy Labels and analyzed the data collection practices of apps according to these labels~\cite{balash2022longitudinal, li2022understandingios}. Through small-scale studies, researchers have discovered that inaccurate APLs can exist due to the developer's knowledge gaps or resource limitations~\cite{li2022understanding}. 

However, a considerable gap remains in comprehensively understanding the developer's interaction with privacy labels, particularly related to Google's enforcement of data safety sections. This interaction includes determining intended privacy practices, overcoming platform challenges, accurately submitting privacy label forms, and updating them over time. Prior research has not accounted for this wide-ranging perspective. In this paper, we address this gap by focusing on Google's Data Safety Section, answering three key research questions:\smallskip

\noindent \textbf{RQ1:} How do developers report their app privacy practices in Google's Data Safety Sections? \smallskip

\noindent \textbf{RQ2:} How have these patterns of reporting practices evolved over the year following the implementation of DSS? \smallskip

\noindent  \textbf{RQ3:} What are the driving factors behind the changes in Google's DSS, the challenges, and the behaviors of app developers? \smallskip
 
To answer these questions, we conduct a large-scale analysis of app privacy labels on the Google Play Store. We periodically scrape the Google Play Store to collect metadata, including permissions and DSS forms, for over \nnumber{2M} apps between June 2022 and May 2023. Subsequently, we engage with 3500 developers of these apps to understand their decision-making process concerning privacy practices, the completion of DSS forms, and their subsequent updates. Our qualitative analysis of developer responses helps answer the third research question by forming an analytical framework to model the developers' experience with DSS.

Our responses to the research questions offer new insights into Google's DSS: 
\begin{itemize} 
\item As of May 2023, privacy labels are only present for 46.8\% of apps on the Google Play Store. Among those apps featuring DSS, we observe patterns of underreporting privacy practices, overreporting practices, and submitting inconsistent DSS forms.

\item Our longitudinal analysis unveils a dynamic landscape for DSS, suggesting that developers are still refining their comprehension and implementation of DSS requirements. About 40\% of the apps updated their DSS at least once over the past year, adding and removing high-level privacy practices and data categories.

\item Our qualitative analysis of developer responses indicates that developers confront challenges when aligning their intended practices with DSS forms. These challenges often lead them to prioritize successfully submitting DSS forms over accurately populating them. \end{itemize}

Finally, we propose novel recommendations to enhance the developer experience with DSS based on our findings. These recommendations include action points for platforms and regulators, such as better educational resources, multilingual support, form simplification, consistent feedback, improved support for third-party libraries, and mechanisms to solicit developer feedback. Lastly, we plan to release a large-scale dataset of the metadata for \nnumber{1.14M} Android apps spanning June 2022 to May 2023.

\section{Background and Related Works}
\label{sec:background}

\begin{figure}[t]
  \centering
  \includegraphics[width=\columnwidth]{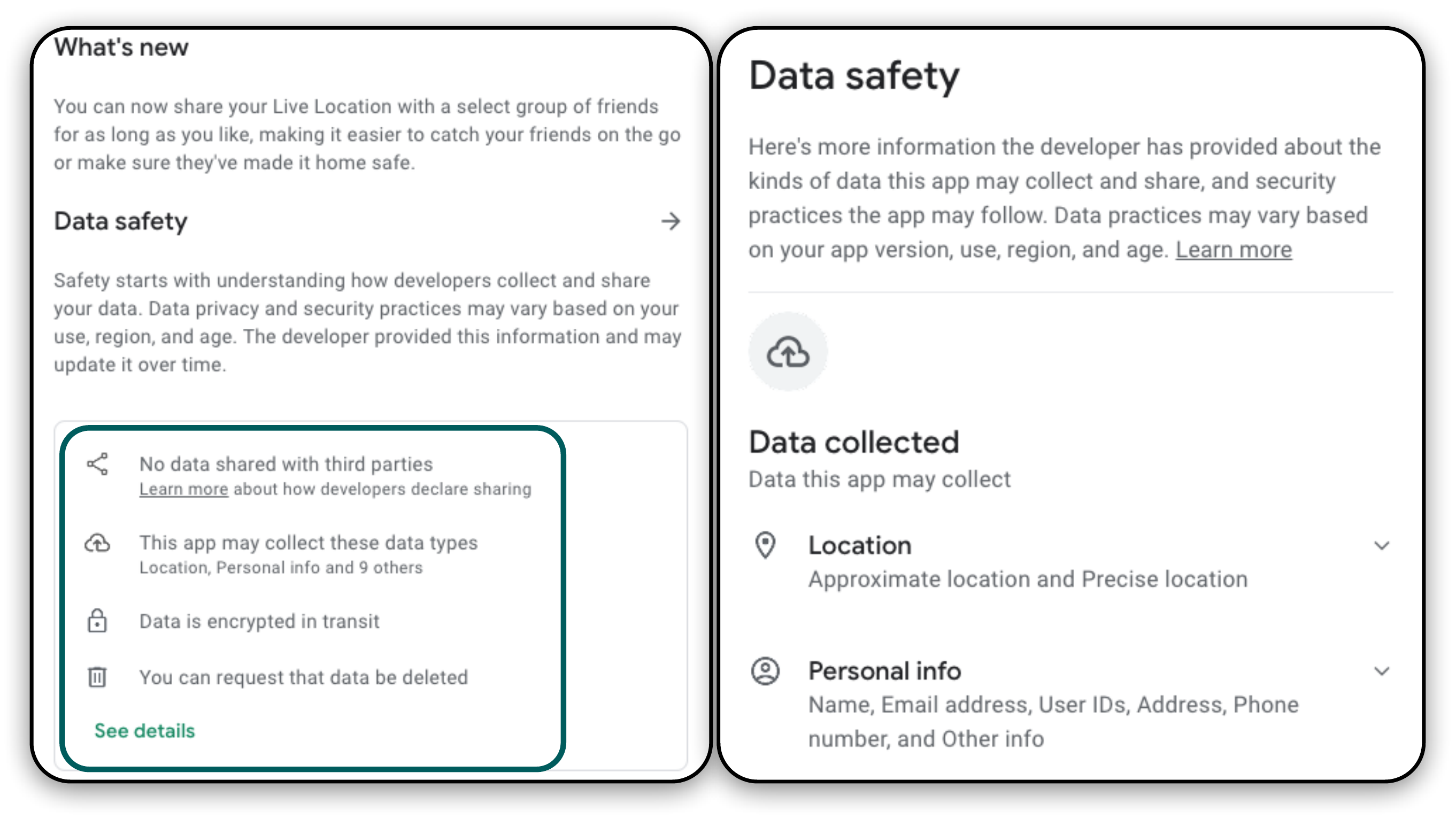}
  \caption{An example of the data safety section of an Android app.}
  \label{fig:pl_example}
\end{figure}

\textbf{Privacy Nutrition Labels.}
Originally introduced by Kelley et al.~\cite{kelley_labels, kelley_study_label}, privacy nutrition labels summarize the privacy practices of websites in a nutrition label format, making them easier to understand. They later designed the ``Privacy Facts'' display to allow the users to consider privacy while installing apps~\cite{kelley2013privacy}. More recently, researchers proposed an Internet of Things (IoT) security and privacy label~\cite{emami2020ask, emami2021privacy} to surface privacy and security information about IoT devices to the users. Researchers have also studied the design and evaluation of privacy notices and labels~\cite{balebako2015impact, kelley_labels, kelley_privacy_app, kelley_study_label, kelley2013privacy, cranor2012necessary,schaub2015design, mcdonald2009comparative, fox2018communicating, cranor2022mobile}.
\smallskip

In 2020, Apple adopted the privacy nutrition labels for the app store and mandated that app developers provide their apps' privacy information in the form of the Apple Privacy Label (APL). Later in 2022, Google required developers to add a Data Safety Section (DSS) to the Google Play Store. Examples of Google's privacy labels are in \Cref{fig:pl_example}. \smallskip

\noindent
\textbf{Google Data Safety Section}
\label{sec:data_safety}
\begin{figure}[t]
  \centering
  \includegraphics[width=\columnwidth]{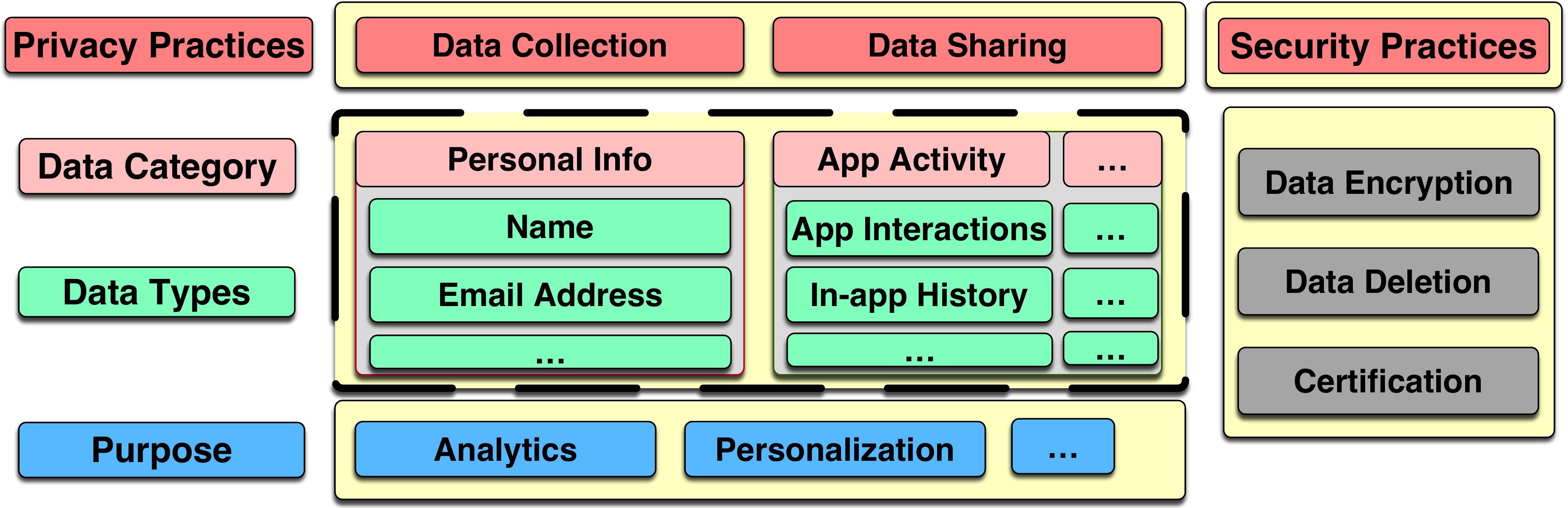}
  \caption{The hierarchy of the Google Data Safety Section, showcasing the various layers and components.}
  \label{fig:google_dss}
\end{figure}
The Data Safety Section (DSS) consists of four levels (\Cref{fig:google_dss}), where the first is high level \textit{Privacy Practices}. The second and third levels consist of \textit{Data Categories} and \textit{Data Types}, and the fourth level consists of \textit{Purpose}.

The first level includes three practices: \textit{Data Collection}, which covers the details about the data that is collected and its intended use; \textit{Data Sharing}, where the developers disclose what data is shared with third parties; and \textit{Security Practices} that covers the data practices related to user choice and data security. \textit{Security Practices} include three tags: \textit{Encrypted in Transit}, \textit{Data Deletion Option}, and \textit{Review against Global Security Standards}.

In the second level, \textit{Data Categories} includes 14 categories such as \textit{App Info} and \textit{Performance and App Activity}. Each \textit{Data Category} can also have \textit{Data Types}, which provide fine-grained information about the data used by the app. For example, \textit{App Activity} includes \textit{App Interactions and Installed App}, as shown in \Cref{fig:google_dss}. The final level of the Data Safety Section consists of \textit{Purposes} that describe the reasons for collecting or sharing the data.\smallskip

\noindent
\textbf{Google App Submission Review.} Google reviews submitted apps to the Play Store to ensure their compliance with its guidelines about design, content, and style~\cite{google_app_review}. As part of these guidelines, Google requires apps to comply with its data safety policies. In particular, all developers must specify the data collected and shared by their app, including the data handled by third-party libraries or SDKs~\cite{google_data_safety}. 

To provide developers with better expectations towards the privacy criteria during the app review process, Google launched the \textit{Checks} service. It gives developers the ability to verify that their apps comply to the data safety policies before submitting the app for review~\cite{google_checks}. The Data Monitoring feature of Google \textit{Checks} monitors multiple channels of data collection and sharing, including with SDKs, via in-app permissions, and to external sites. The service provides developers evidence, such as permissions and network traffic, that the data safety form or the privacy policy is non-compliant. \smallskip

\noindent
\textbf{Usability of Privacy Labels.} Researchers have studied the usability of APLs from both users'~\cite{zhang2022usable} and developers'~\cite{li2022understanding} perspectives. From the developers' perspective, Li et al.~\cite{li2022understanding} interviewed 12 iOS developers and reported that developers err by under-reporting and over-reporting data collection in privacy labels. They further concluded that the label design is confusing for the developers either due to known factors (lack of resources, improper documentation) or unknown factors (preconceptions, knowledge gaps).  Xiao et al.~\cite{xiao2022lalaine} characterize non-compliance of Apple privacy labels by studying data flow to label consistency of 5K iOS apps. They provide insights for improving label design based on their characterization. Researchers also built and evaluated a tool~\cite{gardner2022helping} that helps iOS developers generate privacy labels by identifying data flows through code analysis. \smallskip

\noindent
\textbf{Studies on Privacy Labels.} 
The works most similar to ours perform longitudinal measurement of privacy labels to understand the adoption and evolution of Apple privacy labels over time~\cite{balash2022longitudinal, li2022understanding, scoccia2022empirical}. In particular, Scoccia et al.~\cite{scoccia2022empirical} conducted an empirical study of 17K apps to characterize how sensitive data is collected and shared for iOS apps. They found that free apps collect more sensitive data for tracking purposes. Li et al.~\cite{li2022understanding} and Balash et al.~\cite{balash2022longitudinal} collected weekly snapshots of Apple privacy labels and characterized the privacy practices mentioned in privacy labels for \nnumber{573k} apps. Balash et al.~\cite{balash2022longitudinal} also perform additional correlation analysis with app meta-data like user rating, content rating, and app size. \smallskip

\noindent
\textbf{Our Contributions.} 
In this paper, we investigate the data safety sections of the apps listed on the Google Play Store, observing their evolution over the course of a year. Based on our measurement, we conduct a large-scale study by interacting with more than 3,500 Android app developers, using their apps as case studies. This analysis allows us to model the developers' engagement with the Data Safety Section ecosystem, gaining insight into their challenges, data practices, and the factors influencing their decision-making.

Our results complement prior works by showing that developers on the Apple and Google platforms share common problems when completing the DSS and APL. However, we go beyond prior work by substantiating the underreporting, overreporting, and inconsistencies in data practices; studying the process by which developers interact with the data safety sections; identifying strategies they use to bypass the review process; and revealing the factors that result in changes to the data safety sections over time.

\section{Google Data Safety Dataset}
\label{sec:measurement}
\begin{figure}[t]
  \centering
  \includegraphics[width=\columnwidth]{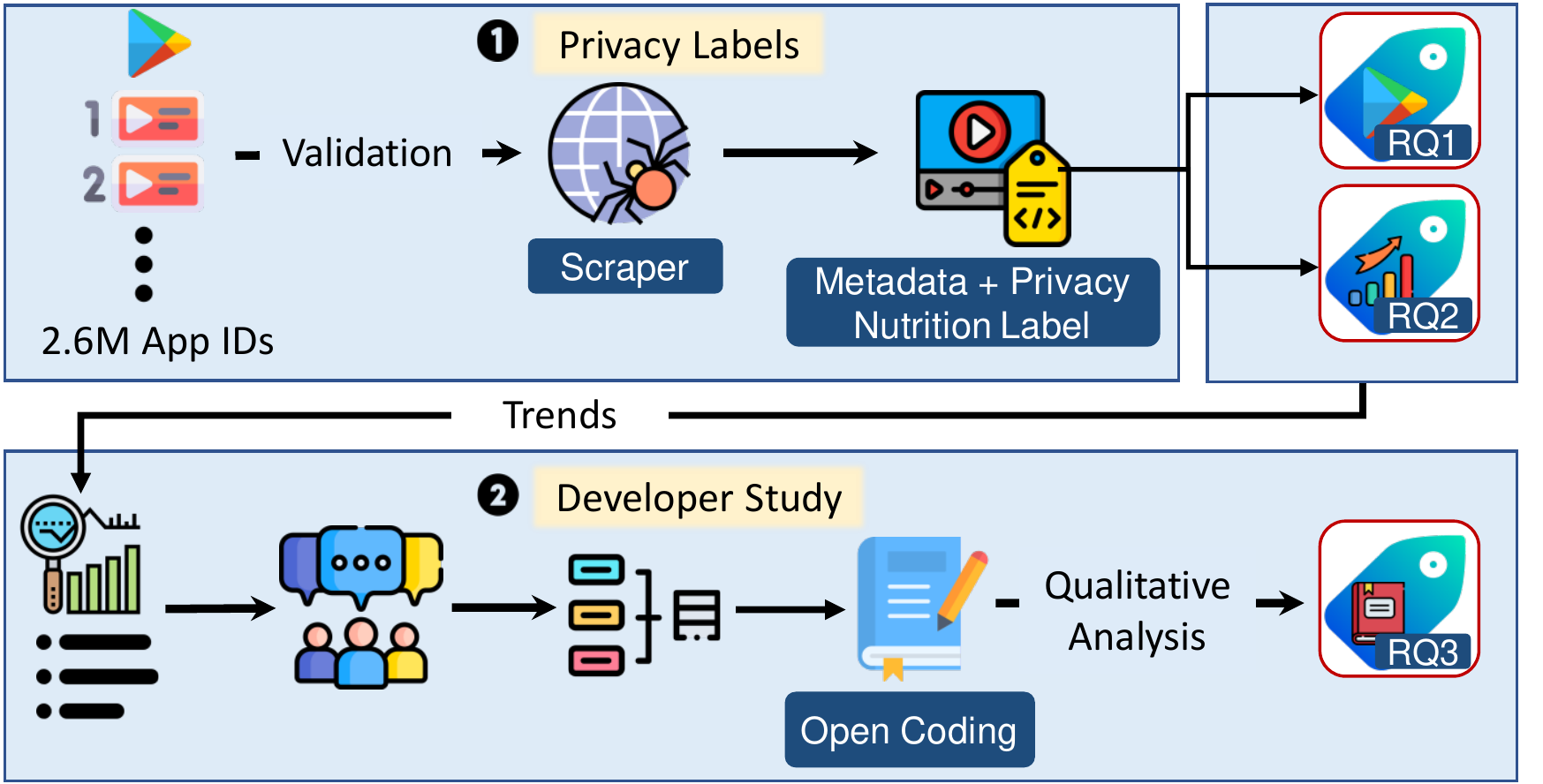}
  \caption{Our data measurement and analysis pipeline consists of scraping the DSS of Android apps to answer the first two research questions and interacting with app developers to answer the third research question.}
  \label{fig:meas_pipeline}
\end{figure}

We curate the Google Data Safety Dataset by scraping the metadata and privacy labels of the apps from Google Play. We contact the developers of some of these apps to understand their process of deciding on privacy practices, filling out the DSS form, and updating the DSS. We show an overview of the analysis pipeline in  \cref{fig:meas_pipeline}.

\paragraph{Dataset Collection}
We took \nnumber{10} snapshots of the Data Safety Sections for \nnumber{2.46M} apps present on the play store between June 20, 2022, and May 31, 2023. We captured eight weekly snapshots from June 20 to Aug 1 and three more snapshots on September 2022, November 2022, and May 2023.  We chose to gather data more frequently around July 20, 2022, the date Google required app developers to complete the Data Safety Section. This allowed us to observe the developers' immediate response to this requirement. The additional three snapshots enabled us to study the developers' long-term reactions. We refer to this dataset with 10 snapshots as \textbf{DSS Dataset}.

We initiated data collection with the apk list provided by Androzoo~\cite{androzoo}. This daily-updated list consists of Android app ids from various sources, including those from the Google Play store. We capture the metadata of each app, including its data safety sections, using the app IDs and a customized version of publicly available google play store scraper library~\texttt{google-play-scraper}~\cite{google-play-scraper}. Across all the snapshots, we observed a total of 2.72M unique apps and 2.17M common apps. For the latest snapshot (May 2023), we retrieved metadata for \nnumber{2.46M} apps, which includes apps with very low download counts. To ensure that our statistical analysis is not skewed by these apps, we filter out apps that have fewer than 1000 downloads resulting in a total of \nnumber{1.1M} apps with \nnumber{539k} having privacy labels.

\paragraph {Dataset Statistics}
Our dataset shows that developers have been slow to add privacy labels to their apps, even after the hard deadlines have passed. Privacy labels are present only for \nnumber{46.8}\% of the apps on the Google play store (as of May 31 2023). We also break down the DSS adoption rate according to app metadata, focusing on the apps'  number of downloads, age rating, and pricing. \smallskip

\noindent \textit{Number of Downloads:} We examine the relationship of the DSS adoption rate and the number of app downloads. In particular, we categorize the apps based on their download numbers on a logarithmic scale, adhering to Google Play Store's binning methodology: (1000+, 5000+, ..., 5B+, 10B+).  We observe a perfect correlation between the adoption rate and the downloads. Starting at \nnumber{38.96\%} adoption rate of apps with 1000+ downloads, the adoption rate increases monotonically reaching \nnumber{100\%} for apps with 10B+ downloads. This observation suggests that developers of highly downloaded apps tend to place more emphasis on the DSS, potentially due to access to greater resources and concern over public perception. 

\noindent  \textit{Age Rating:} The Play Store categorizes apps based on age ratings, including Everyone, Teen, Mature 17+, and Everyone 10+.\footnote{Google also has Adults 18+ rating, but we found less than 200 apps in this category and decided to filter it out for this analysis.} We observed \nnumber{51}\% of apps with the \textit{Mature 17+} rating have a Data Safety Section (DSS), while the fraction of apps with a DSS in the other age ratings ranges from \nnumber{45}\% (Everyone) to \nnumber{54}\% (Everyone 10+). \smallskip

\noindent \textit{App Pricing:} We study the difference in practices based on whether the app is available for free, free with in-app purchases, or paid. We find that \nnumber{65}\% of the paid apps, \nnumber{59}\% of free with in-app purchases, and, for free apps, only \nnumber{44}\% have DSS. We revisit this observation in the developer study where we highlight how developers report the privacy practices of Ad libraries in free apps.

\paragraph {Ethical Considerations}
We collected data only from publicly available web pages and APIs. While our data collection scripts might load Google's servers, we were careful to not abuse these resources. In particular, we added back-off strategies in case of errors and waited for sufficient time before retrying for the failed cases.

\section{Data Practices in Privacy Labels [RQ1]}
\label{sec:google-data-safety}
\begin{figure}[t]         
     \centering
     \includegraphics[width=\columnwidth]{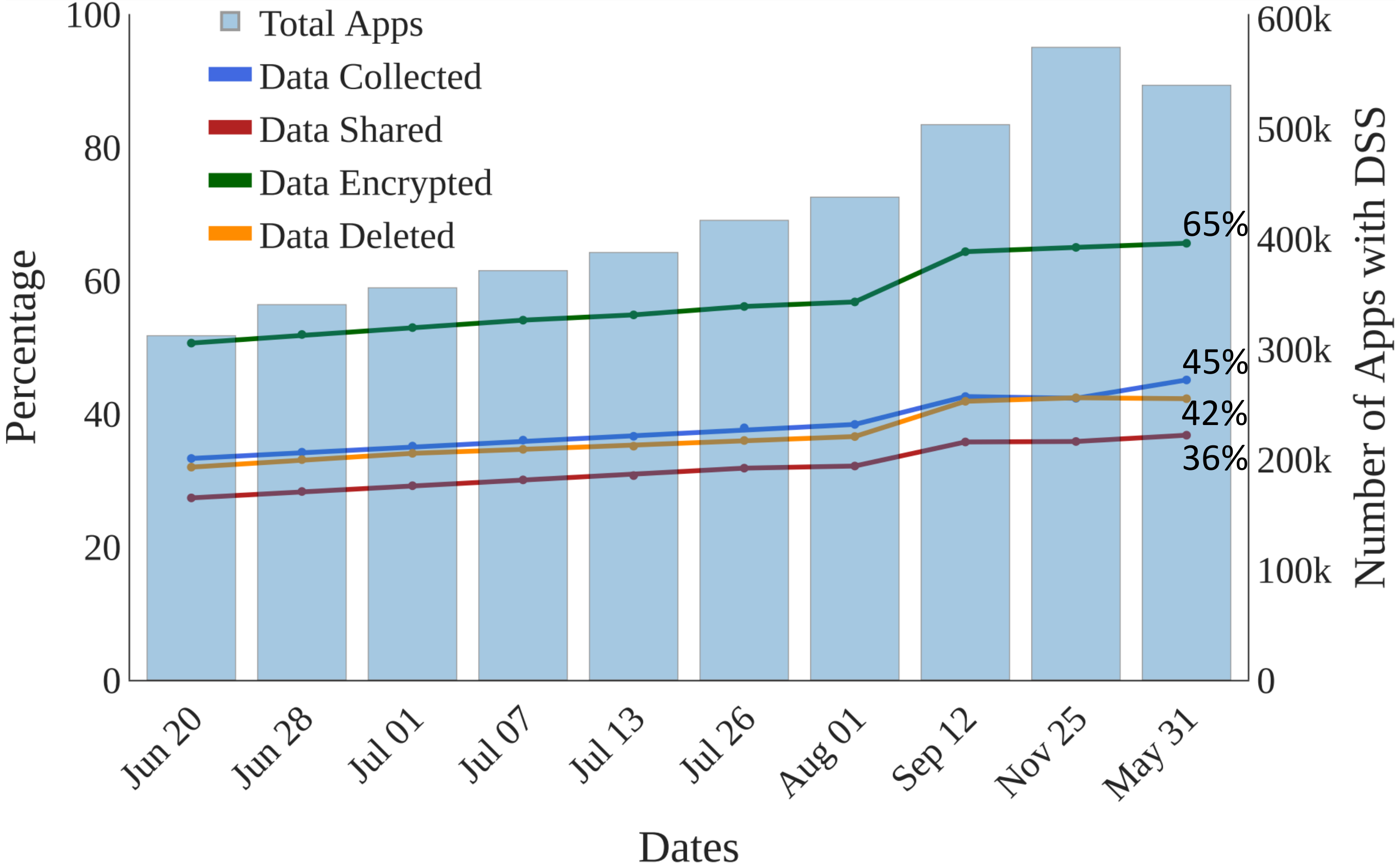}     
    \caption{The evolution of Data Safety Section over the 10 snapshots in our dataset}
     \label{fig:dss_high_level}    
\end{figure}

\begin{figure}[t]
     \centering     
     \includegraphics[width=\columnwidth]{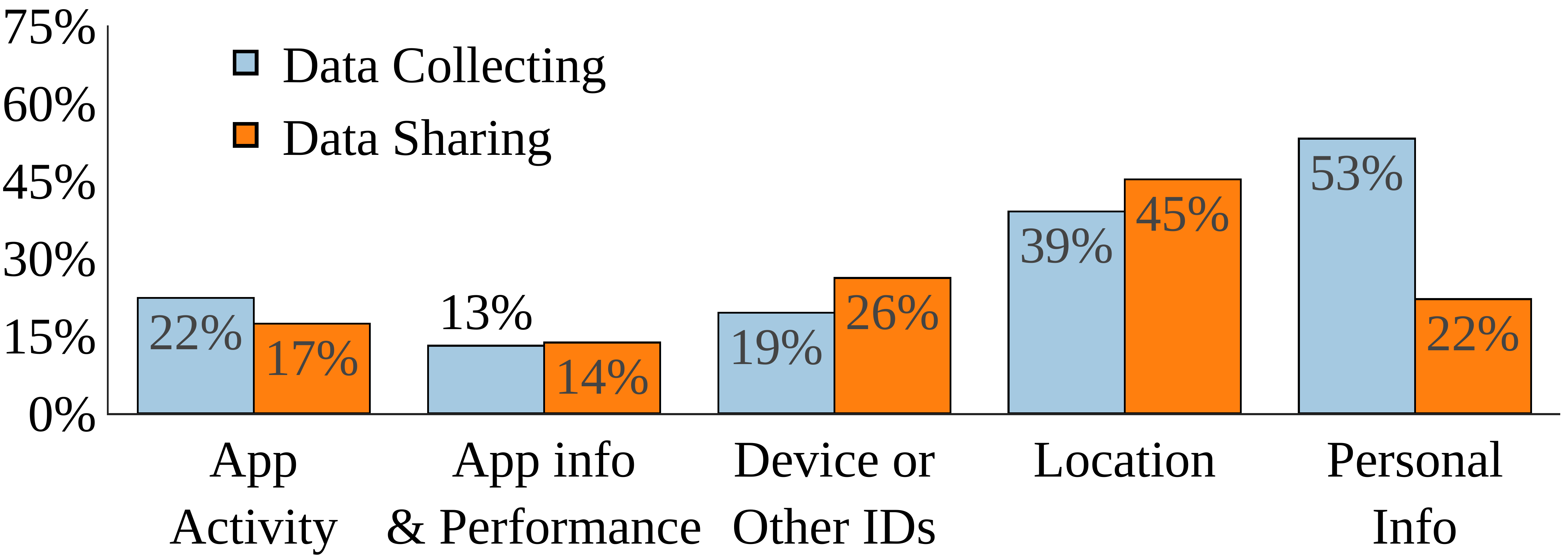}     
    \caption{The distribution of Top-5 data categories for high-level practices}
    \label{fig:dss_datatype}
\end{figure}

We analyze the May 2023 snapshot of the DSS dataset to understand how developers report high-level privacy practices, data types, and purposes. \Cref{fig:dss_high_level} depicts the percentage of apps reporting high-level privacy practices on DSS, with data encryption practices being the most reported. \Cref{fig:dss_datatype} shows how app developers report collection and sharing for the different data categories. Developers report collecting \textit{Location} and \textit{Personal Information} at a higher rate than other data categories, primarily for \textit{App functionality} and \textit{Analytics}. They also report sharing \textit{Location} and \textit{Device Ids} more commonly for \textit{Advertising or Marketing} purposes. The heatmap of \cref{fig:heatmap} (Appendix~\ref{app:figures}) provides a detailed breakdown of the declared purposes for collected and shared data types. 

We also note that apps may declare sharing data without collecting it, as evident from \cref{fig:dss_datatype}. This discrepancy arises from the definition of \textit{Data Collection}, which covers developers retrieving user data from the device using the app~\cite{googledocumentation}. Whereas \textit{Data Sharing} denotes the cases when the data is transferred from the device to a third party. Thus, developers can share data without collecting it if their application employs third-party libraries that directly send data to their servers.

We further analyze this snapshot to understand how developers interact with the DSS forms. Our analysis reveals three key patterns: overreporting privacy practices, underreporting privacy practices, and submitting inconsistent practices.

\paragraph{Underreporting Practices:} Analyzing \Cref{fig:dss_high_level}, we note that only \nnumber{36}\% reported sharing at least one data type, suggesting that the majority of the apps on the play store do not report sharing data. This figure contrasts the findings from prior work~\cite{lin2013understanding,wang2015wukong}, which found that the majority of the apps use at least one third-party library collecting sensitive information~\cite{lin2013understanding}. We confirm this observation by analyzing the 15K most popular apps from our snapshot that report not sharing data with third parties. Specifically, we download the apk file from androzoo and use the LibRadar library ~\cite{androzoo, libradar} to determine whether third-party libraries are used. Given an apk file, LibRadar~\cite{libradar} identifies the third-party libraries and tags them into categories like Advertisement, Analytics, etc. 

Analyzing the apks, we find that 42\% of the analyzed subset of apps used at least one third-party library for advertisement or analytics. This result clearly indicates that developers are underreporting the sharing practices. One possible explanation is that the privacy practices of third-party libraries are often vague and developers find it hard to understand the collection and sharing practices of third-party libraries. We find evidence supporting this explanation in prior work~\cite{balebako2014privacy, li2022understanding} and our developer study (\cref{sec:developer}).

\paragraph{Overreporting Practices:}
Analyzing the purposes for \textit{Data Type} collection, we observe that many apps report a large number of purposes when listing datatypes. We note that out of the 7 possible purposes for collecting data, more than \nnumber{3.5K} apps list 6 or more purposes for every data type they collect, which may indicate that app developers list all purposes out of convenience. For example, \textit{Workplace from Meta} with over 15M+ downloads, lists the same 6 purposes for all the data they collect, like access to \textit{Installed Apps}, \textit{SMS or MMS}, \textit{Music Files}. We also note that while \nnumber{3.5K} is small compared to the dataset, it still has the potential to impact millions of users.

A possible explanation for this observation could be that app developers lack the knowledge required to fill the DSS and choose to select all options. Another possible reason is that they are unaware of the policies of the third-party applications that they use, and take a cautious approach by overreporting. Findings in prior work~\cite{li2022understanding} and from our developer study (\Cref{sec:developer}) align with this observation. 

\paragraph{Inconsistent Practices:}
We observe the developers report practices that are inconsistent with other declared practices or with the app permissions. For example, we find that \nnumber{40\%} of the apps state that they do not collect or share data, but encrypt the data in transit.  We delved deeper into this observation by cross-verifying security practices with apps' network permission requests. \nnumber{59\%} of apps do not request network permissions, yet state that they encrypt data in transit. 
It is not clear why would apps need to encrypt transit data if they are not collecting or sharing data. As \textit{Encryption in transit} has important implications for privacy, we mark this trend and examine it in detail in our developer study in \cref{sec:developer}. 

We also cross-verified the collected and shared data types from the DSS to the app permissions. Several apps report collecting or sharing several data types without even asking for the corresponding permissions. For example, \nnumber{11.5\%} of the apps report collecting or sharing precise location data without obtaining location permissions. Another example is \nnumber{23.7\%} of the apps report collecting or sharing files and documents without the ``Photos/Media/Files'' permissions.

In \cref{sec:developer}, we identify some of the reasons why developers provide inconsistent privacy labels. One reason is developers over-reporting their practices, as indicated above. Other reasons include developers choosing labels by mistake, misunderstanding the label definition, and not updating the labels after updating app features.

\section{Evolution Of Data Safety Section [RQ2]}
\label{sec:evolution}
\begin{figure*}
     \centering
     \begin{subfigure}[b]{0.49\textwidth}
         \centering
         \includegraphics[width=\textwidth]{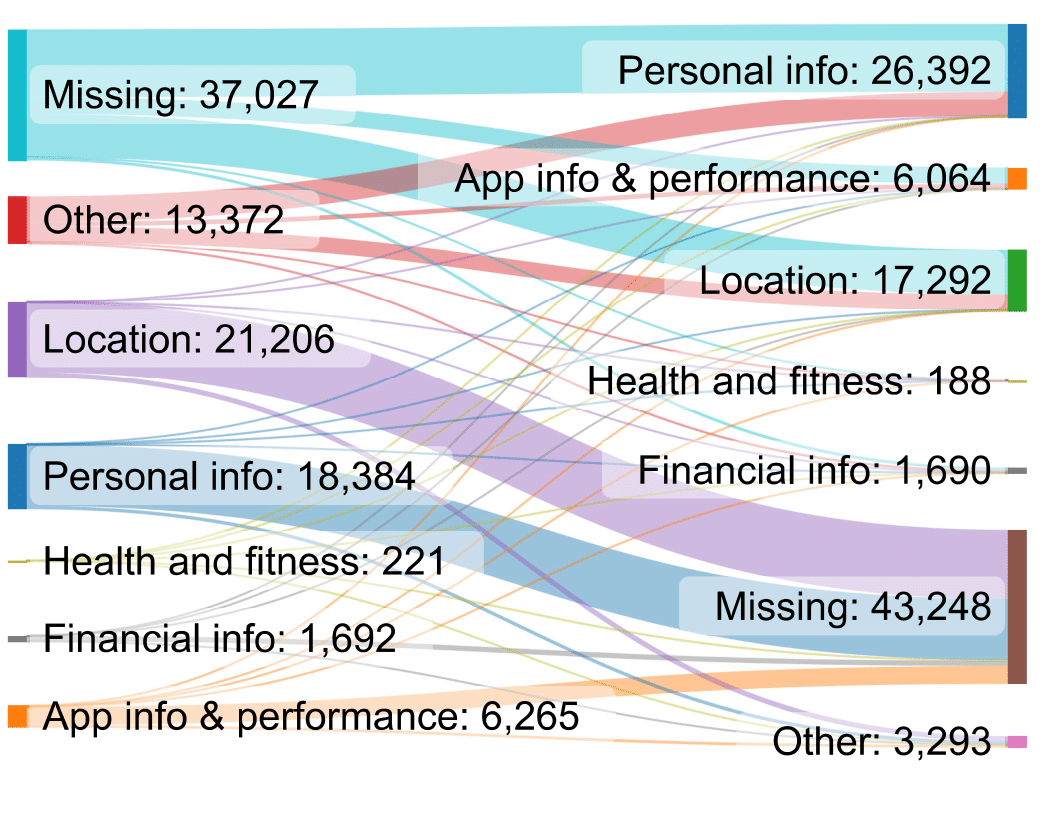}
         \caption{Data Type evolution in Collection}
         \label{fig:evol_coll}
     \end{subfigure}
     \hfill
     \begin{subfigure}[b]{0.49\textwidth}
         \centering
         \includegraphics[width=\textwidth]{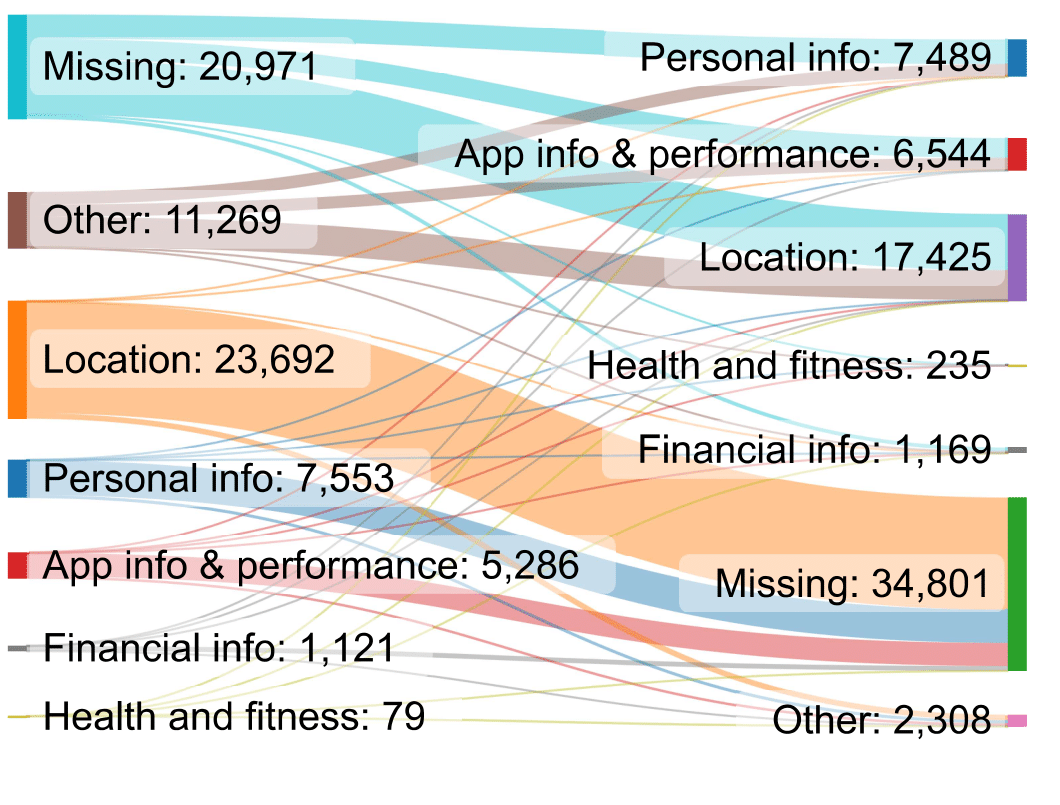}
         \caption{Data Type evolution in Sharing}
         \label{fig:evol_shar}
     \end{subfigure}
     \caption{Change in Data Categories between the first and the final snapshot for (a) Data Collection and (b) Data Sharing. The figure shows that at Data Category level, the reported practices change a lot, indicating that developers change the DSS frequently.}
     \label{fig:sankey}
\end{figure*}
 Next, we conduct a longitudinal analysis of the DSS dataset to understand how the data practices disclosed by developers evolved over time. As described in \Cref{sec:measurement}, our data collection spanned the timeframe before and after the hard deadline set by Google for app developers to comply with the new DSS requirements. This design allows us to understand not only the static state of app data safety disclosures at a given time point but also their evolution as developers navigated this significant policy change. 

 Looking at \cref{fig:dss_high_level}, we observe that during the period spanning June 20, 2022 to May 31, 2023, the number of apps with DSS increased from \nnumber{312K} to \nnumber{539K}. The largest per-day change, of around 9K, happened between June 28, 2022, and July 1st, 2022. Interestingly, we find that from our initial snapshot, 21\% of apps removed their DSS over the course of our data collection. Specifically, \nnumber{67K} updated their play store page to remove the privacy label. For example, \textit{Sport Prediction} with 1M app download had a DSS as of August 1, 2022, but did not have it by Nov 25, 2022.

\begin{figure}[ht]
  \centering
  \includegraphics[width=\columnwidth]{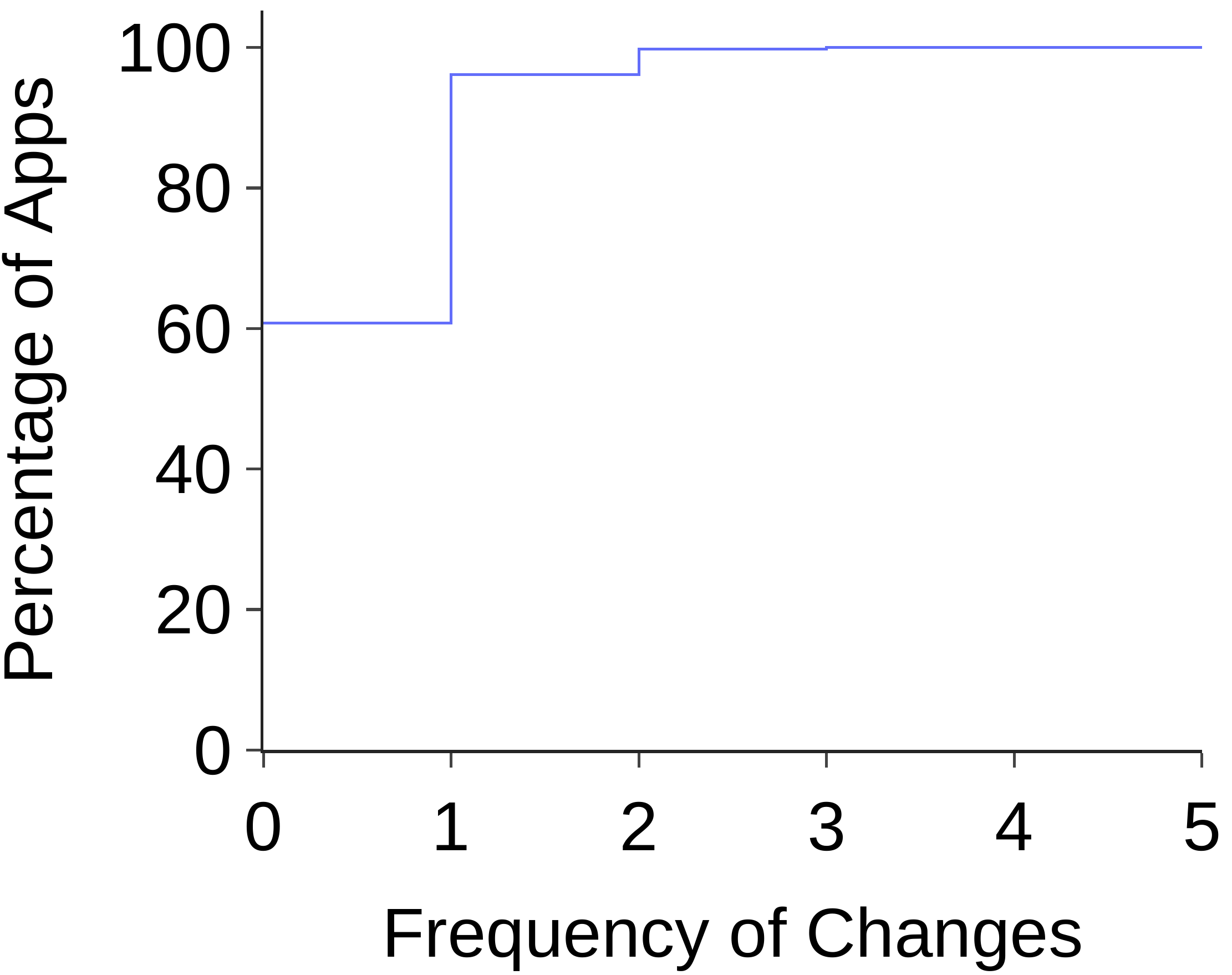}
  \caption{The Cumulative Distribution Function for the frequency of DSS updates made by apps throughout the snapshot timeline.}
  \label{fig:cdf_app}
\end{figure}

\paragraph{Updates in DSS.} We analyze the DSS dataset to understand the frequency of updates in DSS by comparing the first snapshot of the DSS for an app with subsequent snapshots. \Cref{fig:cdf_app} in \Cref{app:cdf} shows the CDF of updates in DSS. We find that 40\% (n=\nnumber{283K}) of the apps updated their DSS at least once, while 4\% (n=\nnumber{27K}) updated it at least twice. Moreover, the frequency of updates is higher in the September and November snapshots. For example, \textit{Adobe Acrobat Reader: Edit PDF}, with over 500M downloads, updated their DSS 3 times between Jun 20, 2022, and May 31, 2023. In \Cref{sec:developer}, we discuss potential reasons for DSS changes. These reasons include app feature updates or the discovery of incorrect DSS.

\paragraph{Evolution of High-Level Practices.} We also analyze the evolution of high-level practices present in the DSS. First, we observe a shift in data collection practices. \nnumber{44K}  applications that initially reported collecting data have updated their DSS to state that they no longer collect data. Conversely, some apps (n=\nnumber{45K}) that initially claimed not to collect data have revised their DSS, admitting to data collection. 

We also observe similar trends in data-sharing practices. Several apps (n=\nnumber{42K}) that initially reported sharing data with third parties later updated their DSS, indicating an end to such sharing. On the other hand, some apps (n=\nnumber{37K}) that did not report data sharing in their initial DSS later added such practices. As changes in collection and sharing practices have implications on user privacy, we highlight these trends and analyze them in detail in our developer study in \Cref{sec:developer}. 

We observed a steady increase in the number of apps reporting \textit{Encryption of Data in Transit}, an aspect crucial for data security. The count rose from 157K on June 20, 2022, to 353K on May 31, 2023, demonstrating a steady progression towards improved data encryption practices (\Cref{fig:dss_high_level}). Similarly, we observe an increase in the number of apps providing \textit{Data Deletion Option} where the count rose from 99K to 227K. 
However, despite the overall increasing trend in \textit{Data Deletion Option} and \textit{Encryption in Transit} practices, a closer examination reveals that some apps are incorporating these practices, while others are withdrawing them. This observation implies that developers are still adjusting their Data Safety Sections.

The ongoing changes in reported data practices, even ten months past Google's deadline for DSS implementation, point to a dynamic landscape, suggesting that developers are still refining their understanding and implementation of DSS requirements. We investigate the challenges that developers face in more depth in our developer study (\cref{sec:challenges}).

\paragraph{Evolution of Collection and Sharing Practices.}
Delving deeper into the evolution at the datatype level, we investigate three \textit{Data Categories} collected most frequently: Location, Personal Identifiers, and App Activity and Performance. \textit{Personal Identifiers} include \textit{Personal Info}, and \textit{Device Id and other identifiers}. We show the change in these \textit{Data Categories} in \Cref{fig:evol_coll}.  We find that a significant number of apps initially reporting collection of user location (n=\nnumber{21K}) and personal information (n=\nnumber{18K}) have revised their DSS to indicate no longer collecting these datatypes. Similarly, around \nnumber{6.2K} apps that were initially collecting app info and performance data have updated their DSS, indicating a halt in this data collection practice. 

Conversely, we also identified apps initially not collecting location data or other specific information, have now started collecting these datatypes. This shows the bi-directional nature of these changes. We find similar trends by analyzing evolution of these data types for \textit{Data Sharing} in \Cref{fig:evol_shar}. 

These findings highlight the evolving nature of data practices which can impact users' trust and privacy. If an app shifts from not collecting to collecting certain datatypes, it may expose users to new privacy risks especially as they will not be notified to this change.  The findings can indicate developers' difficulties or confusion in accurately understanding their apps' data practices as we highlight in \cref{sec:developer}.

\paragraph{Trends in Over-Reporting of Data Practices} In \Cref{sec:google-data-safety}, we observe that developers over-reported the purposes for data collection. Analyzing the DSS dataset, we find a persistent pattern in over-reporting over time, even after ten months. This suggests that a considerable number of developers continue to perceive an environment of low risk or consequence in over-reporting their data practices.

We uncover the potential reasons behind this trend in our developer study. We find that Google's current policy enforcement does not impose penalties for overstating data collection practices, likely due to the limitations inherent in compliance checks, that rely on static and dynamic code analysis. Consequently, some developers may be inclined towards a risk-averse strategy, choosing to over-report to prevent potential policy violations. Although the number of apps following such practices is low (n=4K) in our set, the impact on the wider privacy label ecosystem can be substantial as it could affect users' trust in privacy labels.

\section{Developer Study [RQ3]}
\label{sec:developer}

We conduct a study with Android app developers to understand their perspective when engaging with Data Safety Sections. We identified \textit{interesting} patterns (described in \cref{sec:google-data-safety} and \cref{sec:evolution}) in apps' Data Safety Sections and contacted the app developers to inquire about the factors responsible for the patterns, and the challenges that they face.

\subsection{Methodology}

We first describe the study design and analysis methods. \smallskip

\noindent
\textbf{Study Design.}
We reached out to app developers through emails, probing their experiences and asking specific questions about the privacy labels of their apps. We crafted these questions carefully to stimulate responses. Thus, we contact the developers once via email and analyze their responses to answer our research questions. As such, our study achieves a broader understanding of developer perspectives as compared to prior work~\cite{li2022understanding}.

When emailing developers, we clearly identified ourselves as researchers and stated that we were studying their application and sought information related to their data safety section (\cref{app:dev-study}). Additionally, We refrained from collecting any personal data from the developers, using only the contact information available publicly on the Play Store to contact them. As such, the study has been approved by the Institutional Review Board (IRB) at our institute. \smallskip

\noindent
\textbf{Developer Selection.} Recall that in \Cref{sec:google-data-safety} and \Cref{sec:evolution}, we identified the following three trends.
(A) apps stating that they encrypt data without collecting or sharing data, (B) apps changing their practice from not collecting/sharing data to collecting/sharing data, and (C) apps changing their practices from collecting/sharing data to not collecting/sharing data. Trend (A) points to inconsistency in reported security practice, whereas Trend (B) and (C) have implications of user data privacy.
We identified the apps corresponding to these patterns, sorted them in order of real installs and contacted the top 10,000 developers for each pattern. In response to our emails, we received 3500 responses. To get a clean set for analysis, we filter out automated replies and non-english responses, leaving us with 889 responses.  We then qualitatively analyze these responses to explore how developers describe their privacy practices, the challenges they face while working with the privacy labels and the factors that prompted them to update their data safety section. \smallskip

\begin{figure*}
  \centering
  \includegraphics[width=\textwidth]{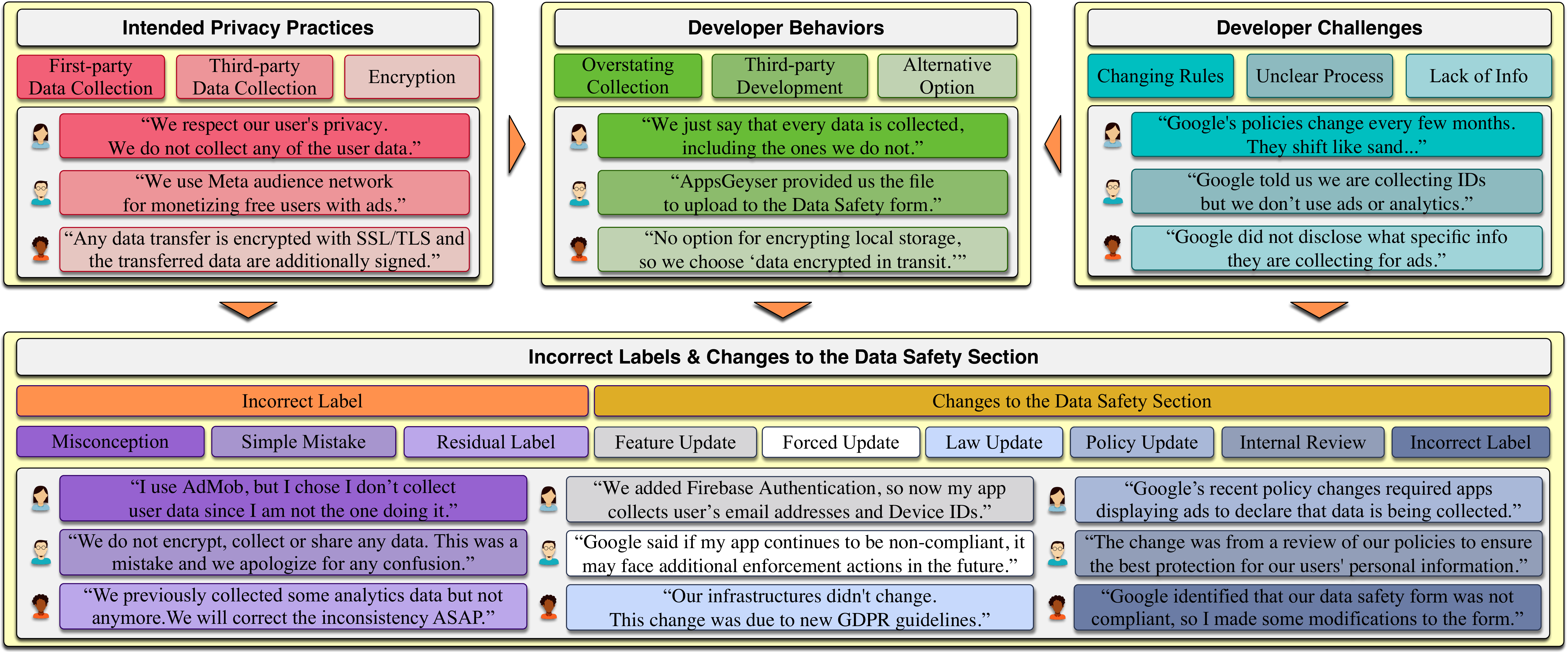}
  \caption{Our analytical framework with the high-level themes, secondary-level themes, and developer quotes. The framework describes the lifecycle of updating the DSS as perceived  by the developers. We do not show all secondary-level themes due to a lack of space.}
  \label{fig:qualitative_study}
\end{figure*}

\noindent
\textbf{Qualitative Analysis Method.} 
In accordance with qualitative research guidelines, we employed a strategy of random sampling and coding of developer responses until data saturation was achieved~\cite{saturation2017}. In our analysis, we reached saturation after coding 225 responses. The coding process began with two authors evaluating an initial set of 50 responses and developing preliminary codes. Next, the research team discussed these codes, clarified differences, and established the initial codebook. The notable differences were in the granularity of the codes and we opted for fine-grained codes as they can be consolidated later. Subsequently, the two authors independently coded a randomly selected subset of 25 responses each time, comparing their codes and iterating on the codebook until they achieved high inter-rater reliability ($\kappa=.87)$) by the 125th response. Using the refined codebook, the two authors continued to code independently until the 225th response, marking the point of data saturation. After conducting a thematic analysis, we model the developers' engagement with the Data Safety Section ecosystem through an analytical framework as depicted in \cref{fig:qualitative_study}. Our codebook is available online.\footnote{\burl{https://osf.io/92sez/?view_only=0b4aa040161a4b259c0a32c7fb3ae82c}} \smallskip

\noindent
\textbf{Findings Overview.} 
\Cref{fig:qualitative_study} describes the lifecycle of updating the DSS as perceived  by the developers. App developers go through several stages while working with Data Safety Section. First, they deduce their intended privacy practices (\Cref{sec:privacy_practices_dev}) based on their app implementation. While attempting to fill the DSS form, they face challenges (\Cref{sec:challenges}) mapping their intended practices to the form. The challenges result in strategies (\Cref{sec:developer_behavior}) to initially fill out the form and get it accepted. After submitting the initial form, there might be inaccuracies highlighted to them either via Google or through their internal review system. Finally, they update the form (\Cref{sec:change_dsc}). In the subsequent sections, we present the findings of our qualitative analysis, unpacking the experiences and behaviors of app developers as they go through this process of uploading their Data Safety Sections.

\subsection{Intended Privacy Practices of Developers}
\label{sec:privacy_practices_dev}
We first discuss developers' intended data collection and sharing practices. Our thematic analysis reveals three subthemes within the privacy practices of the apps: third-party data collection, encryption, and first-party practices.

\smallskip
\noindent
\textbf{Third-party Data Collection:} We find that developers often reported involving third parties for various purposes. The main purposes were ads and analytics. We also find here that developers tend to use these libraries primarily for ad revenue. For example, one developer expressed: ``\textit{... to make profit from my DJ app, I use third party advertising, such as Google AdMob and AppLovin. They (third party ads) are collecting users data to showing perfect ads to my DJ users}.'' This is consistent with our findings from \cref{sec:google-data-safety} where we find that \textit{Ads} and \textit{Analytics} are the top two purposes for data sharing.

\begin{figure}[t]
  \centering
  \includegraphics[width=\columnwidth]{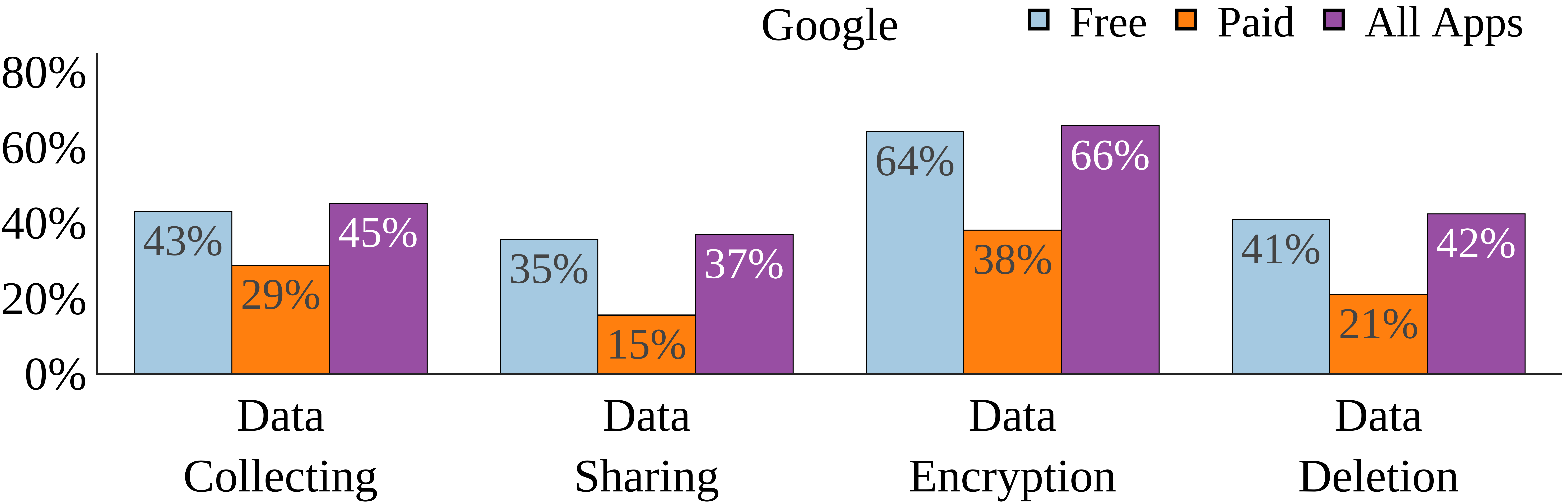}
  \caption{The distribution of high-level data practices based on price. We note that the fraction of paid apps that collect and share data is lower than free apps.}
  \label{fig:price_high_level_app}
\end{figure}

However, we observe developers do not tend to share data when they employ in-app purchases. For example, when asked about changes in their DSS, one developer noted: \textit{``In the last update, I removed all the ads of the app, so I receive money from the few premium subscriptions. The Google AdMob SDK collected all that data, so now it doesn't get any type of information from the user.''} This suggests developers of paid apps tend not to share data possibly as they do not rely on ads for revenue. In \Cref{sec:measurement}, we observed that the fraction of apps with DSS in paid apps is higher than in free apps. Analyzing the \textit{Data Sharing} practices of paid apps (\Cref{fig:price_high_level_app}), 
we indeed find that a lower number of paid apps report sharing data than free apps. \smallskip

\noindent
\textbf{Encryption:} We find that developers often secured data during transmission. One developer reported--``\textit{we currently utilize the IPsec protocol to ensure secure transmission of data. IPsec is a widely adopted industry standard for VPNs and provides robust encryption, authentication, and integrity protection for transmitted data}.'' Some developers also assumed that third-party libraries will provide encryption, although this might raise questions about their direct control over data security. For example, one developer mentioned, ``\textit{... I use Google API like Drive, YouTube, AdMob and Firebase. They do collect user data I guess and they state that data is encrypted}''. \smallskip

\noindent
\textbf{First-party Practices:} In our analysis, developers also conveyed their own data practices. Many developers reported not intending to collect (n=60\%) or share data (n=48\%) themselves. As one developer stated, ``\textit{... we do not collect any personal data or share any personal data. Our app only collects anonymous analytics events (button clicks and screen impressions/load) and app crashes via Google’s Firebase framework so that we can fix user issues and crashes}.'' Others asserted not integrating any ads or analytics services, while some stressed the importance of obtaining user consent before collecting data, underlining their commitment to respecting user privacy.

\subsection{Challenges Faced by Developers}
\label{sec:challenges}
\begin{figure}
  \centering
  \includegraphics[width=\columnwidth]{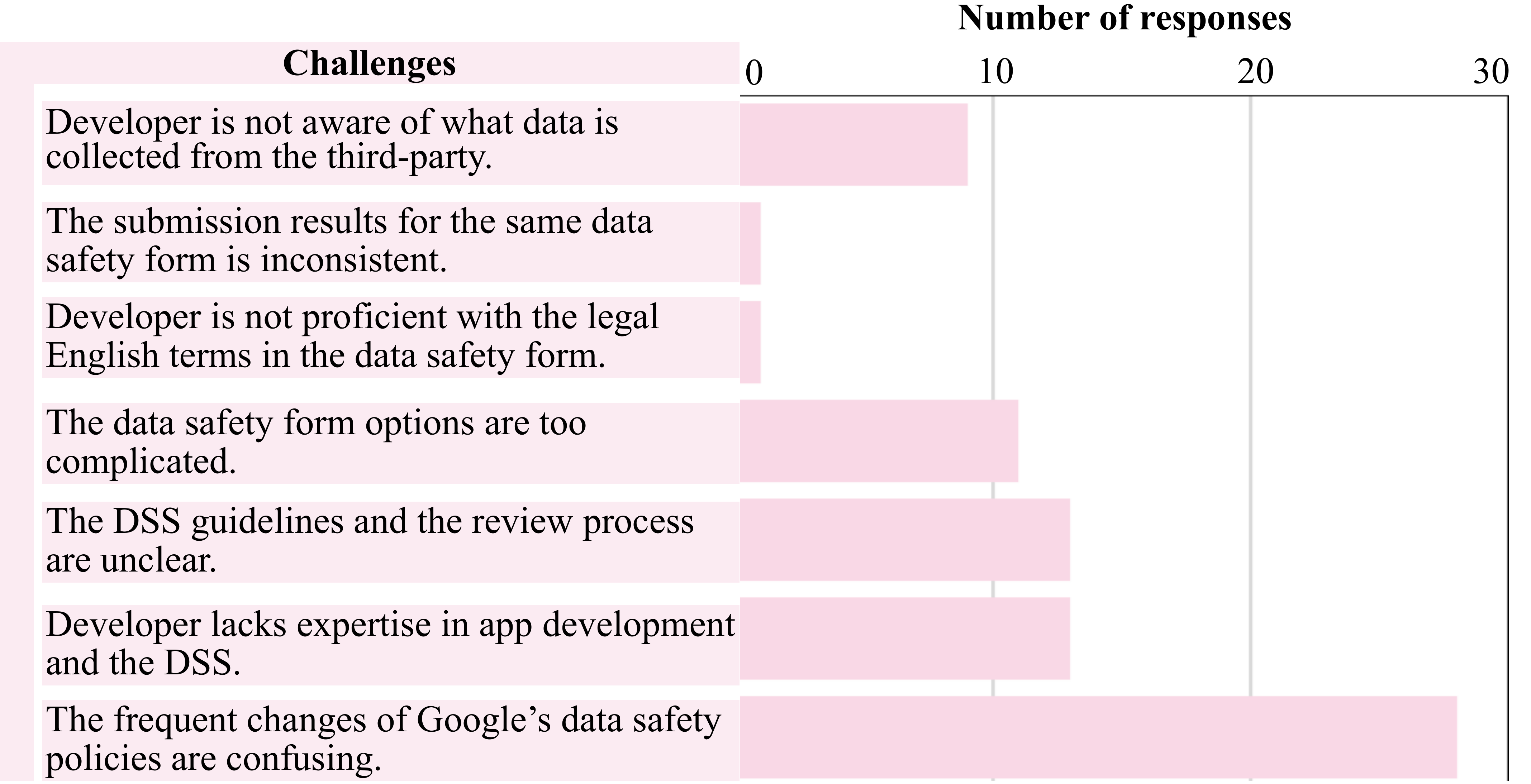}
  \caption{The frequencies of seven subthemes of developer's challenges regarding the Google Data Safety Section, as explained in Section \ref{sec:challenges}.}
  \label{fig:challenge_freq}
\end{figure}

Next, we uncover the challenges that the developers face while mapping their intended privacy practices to DSS forms. 
These challenges are categorized into seven sub-themes with \Cref{fig:challenge_freq} depicting the frequency of each sub-theme.
\smallskip

\noindent
\textbf{Lack of Information About Third-Party Libraries:} Developers indicated uncertainty about the privacy practices of the third-party libraries used in their apps. One developer noted: \textit{``I currently face a challenge in that I am unsure of what specific APIs are included in the app code provided by Appsgeyser.com that may be contributing to this issue of non-compliance, as said by Google. Additionally, I have been unable to find a reliable source of information on this matter.''} Notably, in the play console, Google shows the permissions required by popular libraries, as well as flags problematic SDKs, however, there is no help regarding the collection practices of these libraries. \smallskip

\noindent
\textbf{Inconsistency:} Some developers reported inconsistency in the acceptance of their data safety form. A developer struggling with the review process noted: \textit{``... we submit the same form over and over again and often times Google rejects our answers with no or at most a vague explanation. Eventually, Google accepts it.''} Such inconsistency can lead to confusion about what is required, and result in frequent revisions, and inaccurate DSS. \smallskip

\noindent
\textbf{Lower English Proficiency:} Non-native English speakers cited difficulties understanding the data safety policies. For example, a developer, whose native language is not English, was confused with an English word: \textit{``... our administrator whose first language isn’t English did not seem to understand the meaning of 'ephemerally' and ticked 'No, this collected data is not processed ephemerally'. So even though we declared the data collected/shared, Google play did not disclose this on the app’s store listing.''} \smallskip

\noindent
\textbf{Complicated:} Developers also complained that options provided in DSS form are complicated. For instance, a developer made mistakes in the data safety form because there were simply too many options, whereas another was confused with the complicated description of the DSS option. This complexity can lead to errors or misunderstandings when filling out the form. \smallskip

\noindent
\textbf{Unclear Process:} Our analysis also revealed that developers are confused regarding Google's review process for DSS, finding the process unclear and the explanations vague.  For instance, a developer noted \textit{``Google at one time prompted me that my app collects data of which we knew nothing about, the data they spoke about is the 'Devices and Other IDs Data' and that it's a compulsory data been collected by most apps on Play Store even if they do not collect user data. We don't see or collect any user data from my users and our app doesn't not even use firebase, one signal or ads.''} \smallskip

\noindent
\textbf{Lack of Expertise:} Developers also acknowledged difficulty arising from a lack of expertise in both app development and data safety policies. For example, one developer noted--\textit{``... due to my limited experience, I made this app with the help of third party app developer-AppsGeyser. I am not sure of how I should represent the third-party APIs in the app code and I have been unable to find a reliable source of information on this matter..''}. Another developer noted that they do not understand half of the DSS policies and so they just keep submitting answers in hopes of finding the right configuration. This gap in knowledge can happens due to vague definitions in data safety policies. Moreover, the lack of information can lead to developers having an incomplete understanding of the privacy practices, resulting in inaccurate representation of these practices in the DSS. \smallskip

\noindent
\textbf{Changing Rules:} Developers reported confusion stemming from changing DSS policies within the Google Play Store. They mentioned that the constant evolution of these rules made it difficult for them to ensure compliance. As one developer said: \textit{''... so we haven't changed anything in <app\_name> in almost a decade. What has changed, and seems to keep changing every few months is Google\'s privacy policies. They are difficult to understand and they shift like sand ...''}. This indicates that developers have difficulty keeping up with the dynamic landscape of rules and regulations regarding DSS. \smallskip

These challenges highlight the need for more developer support in areas such as policy comprehension, form simplification, and the handling of third-party libraries. Addressing these issues could significantly aid developers in enhancing data safety and privacy in their applications.

\subsection{Developer Behaviors and Strategies in DSS Submission}
\label{sec:developer_behavior}
\begin{figure}
  \centering
  \includegraphics[width=\columnwidth]{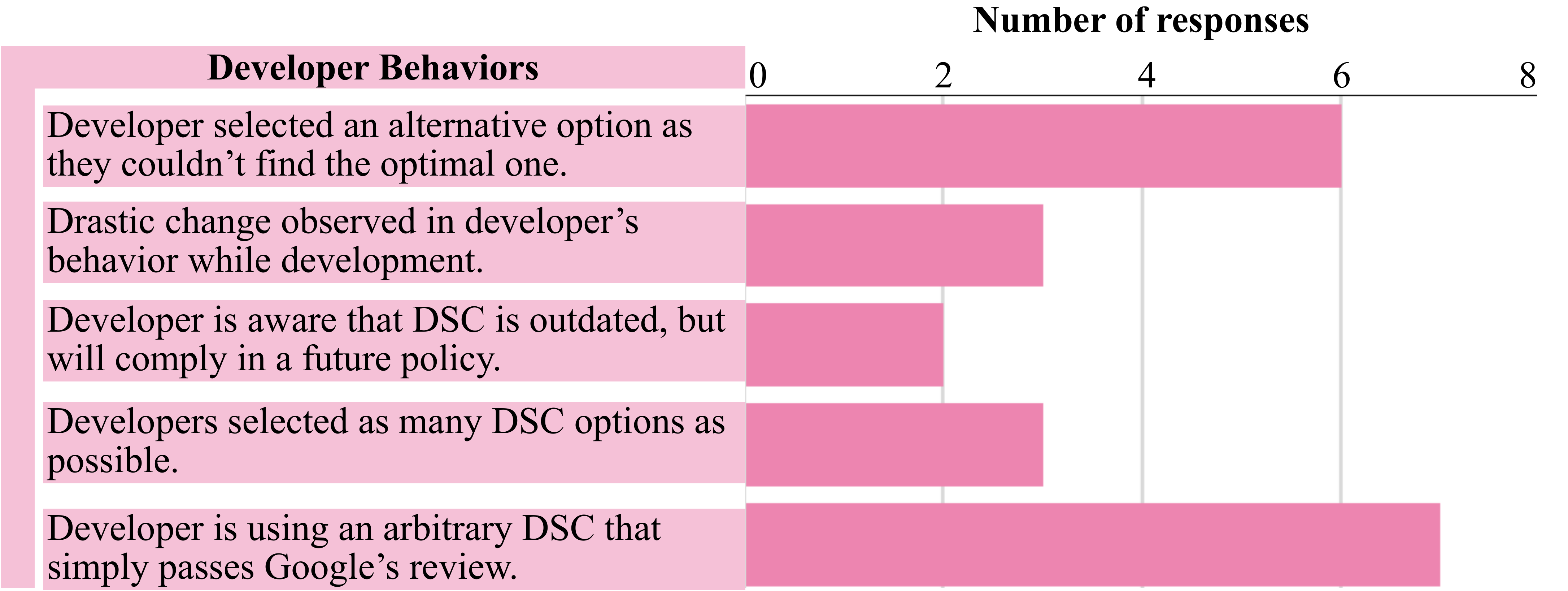}
  \caption{The frequencies of five subthemes of developer's behaviors regarding the Google Data Safety Section, as explained in Section \ref{sec:developer_behavior}.}
  \label{fig:dev_behavior_freq}
\end{figure}

We now discuss the various strategies adopted by the developers to navigate the challenges they face while submitting the DSS form. These behaviors and strategies are classified into five sub-themes. \Cref{fig:dev_behavior_freq} shows the frequency of each sub-theme in the coded responses.

\smallskip

\noindent
\textbf{Alternative Option:} When developers were unable to find the optimal data safety card option for their app, they resorted to choosing an alternative one. For instance, as said by one developer--``\textit{... we do not collect any data, everything user write on our app will be stored on their device. There wasn't any other option to select except 'data is in encrypted in transit ...}'' . This highlights the potential need for more comprehensive or flexible options within the form.

\noindent
\textbf{Change in Developer Behavior:} Our analysis also captured developers altering their behavior based on their experience while interacting with DSS. For instance, one developer that stopped the development of their app noted: \textit{``... considering all the hassle I had been going through, it has been determined that it would not be practical to continue updating these apps on the Google Play Store, given the minimal profit generated and the effort required to maintain compliance.''} This highlights the scenario where the complex policies of DSS can potentially discourage developers to update their apps, hurting the users as they might be forced to work with outdated versions of the apps.\smallskip

\noindent
\textbf{Future Policy:} Some developers acknowledged that they intentionally had an incorrect DSS to increase flexibility, and in case of future app updates. For example, one developer responded: \textit{``It is true that we do not share or collect any kind of personal data. It means that if we intend to do so in the future then we will be encrypting the data. At the moment we are not sharing or collecting personal data.''} This might convey incorrect information to the users. Conversely, if the developer does not indicate that they collect data, but utilize the same flexibility, it could pose a risk to users.\smallskip

\noindent
\textbf{Overstating Collection:} Some developers indicated that Google only checks for the data that is collected but not mentioned as collected but not the other way. As a result, they select all options for collection in the DSS form. For example, one developer noted --\textit{``... it is safe to declare all possible data collected in Data safety section so, in our apps, even the information (e.g. phone numbers) that is not collected is mentioned as collected.''} We also observed similar over-reporting of practices in \Cref{sec:google-data-safety} and \Cref{sec:evolution}. This observation highlights an issue with Google's compliance analysis tools.\smallskip

\noindent
\textbf{Form That Works:} Some developers reported using a data safety form that had previously passed Google's review, regardless of whether it fully aligned with the current data handling practices of the app. For example, one developer noted --\textit{``... my app did not collect any data. but in the data safety section, I wrote that I collect data because other than that Google would not publish my app.''} This approach suggests a pragmatic yet concerning response by the developer as it may result in inaccurate DSS. \smallskip

 The findings highlight that developers attempt to circumvent the review process either by over-reporting their privacy practices or by trying random configurations until their DSS is accepted. Combined with the challenges (\Cref{sec:challenges}) that the developers face, this indicates that the DSS submission tool might not be usable by the developers.

\subsection{Inaccurate Labels and Driving Factors behind Changes in Data Safety Section}
\label{sec:change_dsc}
As the developers navigate the complex challenges, they may submit inaccurate labels. We used the developers' intended practices from \Cref{sec:privacy_practices_dev} and compared them with the reported practices in DSS. Additionally, our qualitative analysis revealed several key factors that influence developers' decisions to update their applications' Data Safety Sections (DSSs).\smallskip

\subsubsection{Inaccurate Labels and Type of Inaccuracies}
\label{sec:incorrect_labels}

We conducted a comparative analysis between the reported practices (\textit{Data Collection} and \textit{Data Sharing}) in DSS and their intended practices from the emails. We find a notable discrepancy; 41\% of developers who stated in their emails not to engage in data collection were nonetheless reported as collecting data in their DSS. A similar inconsistency (42\%) was observed for data sharing. For instance, the developer of an app with over 100M downloads stated in an email, ``\textit{In Google Play Console we pointed that we collect/share data because of Ad network SDKs integrated into the game. Not the app itself, but ad SDKs do collecting or sharing.}'', but their DSS did not report any data collection or sharing. This discrepancy highlights the fact that the DSS sections might not be thoroughly monitored by Google and reiterates the need for more effective mechanisms for accurate reporting of privacy practices. Next, our qualitative analysis also reveals factors responsible for inaccuracies. \smallskip

\noindent
\textbf{Misconception:} Some developers demonstrated misunderstandings of the data safety policies, which resulted in inaccurate DSS.  Such developers have the misconception that they do not have to disclose practices for third-party libraries even when they are using third-party SDKs. For example, one developer noted --\textit{``... we do not collect any personal data or share any personal data. Our app only collects anonymous analytics info through Google Firebase. Hence, we chose that we do not collect data.''} 

Recall the \textit{interesting} trend that we observed in \Cref{sec:google-data-safety} where we noted that many apps state they do not collect or share data, but encrypt data in transit. We find that this could be happening due to misconceptions about using third-party libraries. For example, one developer noted -- \textit{``I use ADMOB for banner ads at the bottom of my app, but Google collects Device ADID. I checked that data is encrypted for that part.''} The developers report \textit{Encryption}, but due to the misconception, do not report \textit{Data Shared}. 

\smallskip

\noindent
\textbf{Simple Mistake:} Some Developers also acknowledged that they made a mistake and conveyed their plans to update their DSS during the communication with the authors. For example, one developer mentioned --\textit{``previously, the process of filling out the Data safety section was interrupted and uncompleted due to one of the checkboxes left unchecked.''}

\smallskip

\noindent
\textbf{Residual Label:} Some developers failed to update their DSS after making feature updates. The developers conveyed their plans to update their forms, but this trend can potentially be detrimental to the DSS ecosystem, especially if the update adds \textit{Data Collection} or \textit{Data Sharing} flows.

\subsubsection{Driving Factors for Change in DSS}
After submitting the initial forms, developers may need to update their DSS. We now discuss the various factors responsible for the change in DSS. Our analysis revealed six distinct sub-themes for these factors. \smallskip

\noindent
\textbf{Feature Update.} One significant trigger for updating DSSs was the introduction of new features. Most commonly, developers updated their DSSs due to the addition of advertisement or analytics SDKs into their app. For example, one developer notes: \textit{"... In the initial version there were no ads. But now I have put ads in the app for which I had to change my policy as well."}\smallskip

\noindent
\textbf{Law Update.} Changes in legal and regulatory frameworks also compelled developers to modify their DSSs. For example, one developer noted -- \textit{``Our methods and infrastructures didn't change. This change was made in light of new GDPR guidelines that are in effect right now.''}\smallskip

\noindent
\textbf{Policy Update.} Changes imposed by the platform (like the Play Store), required adjustments to the DSS. For instance, one developer using a third-party ad service mentioned that: \textit{"We don't collect or transmit any user data. However, some of our apps use Google's AdMob for ads. As per new directions from Google, we are required to reproduce Google data policies related to AdMob even if our app does not collect any data."} \smallskip

\noindent
\textbf{Forced Update:} We also observed instances where developers were compelled to update their DSS to avoid having their apps removed from the Play Store. For example, one developer noted -- \textit{``We detected user data transmitted off device that you have not disclosed in your app’s Data safety form as user data collected. If your app continues to be non-compliant after August 22, 2022, your app updates will be rejected and your app may face additional enforcement actions in the future.''} This showcases the influence of platform policies on developers' decision-making processes.

\smallskip

\noindent
\textbf{Internal Review.} The results of internal reviews also surfaced as a factor for DSS updates. Specifically, a developer mentions that the updated DSS guidelines provided by Google enabled them to more accurately portray how they handle data. They also emphasize that: \textit{"... the added features on the stores rightfully illuminated some areas that were unknown to users, which made a lot of apps appear to handle their data differently than they actually are."} \smallskip

\noindent
\textbf{Incorrect Label.} Finally, developers noted that
the discovery of incorrect labels results in DSS updates.
 As noted by a developer--\textit{"... Although I still do not collect any data, I was essentially required by Google some time ago to say that I collect data on behalf of the advertising services in order to remain compliant with Google's policy ..."}--most of the developers fixed the DSSs after they realized that they chose the incorrect data safety card option due to their confusion after being formally notified by Google. 

This analysis underscores the complex and multifaceted nature of developers' decision-making processes around updating DSSs, influenced by internal practices, external regulations, and the dynamics of app development.

\section{Discussion and Recommendations}
\label{sec:discussion}
Our research has highlighted the complex landscape of developers' experiences and challenges with Data Safety Sections. We follow on our findings by presenting a set of recommendations aimed at improving the developer experience with DSS. Then, we discuss the roles of the platforms and regulators in addressing the deficiencies in DSS. Finally, we list the limitations of our methods. \smallskip

\begin{table}
\begin{tabularx}{\columnwidth}{X}
\toprule 
\textbf{Recommendations} \\
\toprule
\textbf{1. Enhance Educational Resources:} Educational resources should be enhanced to provide developers with real-world examples, interactive tutorials, and guidelines.\\
\midrule
\textbf{2. Provide Multilingual Support:} Resources should be available to developers in multiple languages \\
\midrule
\textbf{3. Simplify Data Safety Forms:} Data safety forms should be simplified to improve comprehension.\\
\midrule
\textbf{4. Consistent Feedback from the Review Process:} Developers should be given clear and transparent feedback on the approval or rejection of their data safety forms.\\
\midrule
\textbf{5. Improve Support for Third-Party Library Data Practices:} Platform providers and third-party library developers should ensure transparency about their data practices.\\
\midrule
\textbf{6. Regular Consultations and Feedback Mechanisms:} Consult with developers to ensure continued relevance of privacy labels; allows developers to share their feedback.\\
\bottomrule
\end{tabularx}
\caption{Summary of the recommendations.}
\label{tab:recom}
\end{table}

\subsection{Recommendations}
We provide a list of recommendations to improve the developer experience with the DSS section in Google Play. These recommendations, summarized in \Cref{tab:recom}, go beyond those presented in prior work~\cite{li2022understanding}. \smallskip

\noindent
\textbf{Enhance Educational Resources.}
An important need emerged for enhancing educational resources surrounding data safety policies and the filling of data safety forms. Such educational resources include detailed explanations about data collection, storage, and sharing practices that are expected to be reported in the Data Safety Section. Providing developers with real-world examples, interactive tutorials, and guidelines to navigate data safety requirements can go a long way in reducing misunderstandings and ensuring proper compliance. \smallskip

\noindent
\textbf{Provide Multilingual Support.}
The global landscape of app development calls for the need to support diverse languages. Developers across the globe should be able to access, understand, and interpret guidelines without language acting as a barrier. Thus, providing guidelines, forms, and supporting services in multiple languages is crucial. Platform providers could consider deploying multilingual support teams and translation services to cater to this diverse community. \smallskip

\noindent
\textbf{Simplify Data Safety Forms.}
Several developers have reported difficulties in understanding the complexity of data safety forms (\cref{sec:challenges}, leading to inaccuracies in reporting. Addressing this concern would involve simplifying the forms to ensure they are easy to understand and complete. This could be achieved by using accessible language, clear terminologies, and unambiguous options. Additionally, redesigning the form layout to enhance readability and ease of use might also prove beneficial. 

\noindent
\textbf{Consistent Feedback from the Review Process.}
In \cref{sec:challenges}, we find that developers have little confidence in the approval process. As a result, developers resort to filling the form to just receive the approval, without the form actually being accurate. Developers expressed frustration as to understanding why a particular form is accepted or rejected. Developers need to have confidence in a consistent and transparent review process. The approval or rejection of data safety forms should be communicated clearly with an explanation of the reasons. This will not only help developers improve their subsequent submissions but also reduce confusion and frustration. \smallskip

\noindent
\textbf{Improve Support for Third-Party Library Data Practices.}
The increasing use of third-party libraries poses challenges for developers to accurately represent data practices in their apps. To mitigate this challenge, platform providers should demand transparency from third-party providers about their data practices. Additionally, developing tools or mechanisms that help developers to track and represent these practices in their data safety forms would be beneficial. An existing effort from Google is the \textit{Google Play SDK Index}, which aims to provide developers transparent information on all the third-party SDKs that is usable in Android app development \cite{google_play_sdk_index}. While the index informs the users of the Android OS permissions these services are requesting, it still fails to provide information on the specific category of data being collected.\smallskip

\noindent
\textbf{Regular Consultations With Developers and Feedback Mechanisms.}
To ensure the continued relevance and effectiveness of privacy labels, regular consultations with developers should be conducted. This would provide a platform for developers to share their experiences, voice their concerns, and give feedback on existing processes. Such feedback mechanisms could provide invaluable insights for platform providers to understand evolving challenges and adapt their policies and support systems accordingly.\smallskip

\subsection{Discussion}

\noindent
\textbf{Impact of Industry Intervention.} Google's recent introduction of the \textit{Checks} service~\cite{google_checks}, providesing paid compliance analysis to developers, adds to the dynamics of the privacy labels. From the developers' perspective, popular apps with resources might find Checks to be an invaluable tool, simplifying the task of policy compliance and mitigating the likelihood of errors or misunderstandings. However, the service's cost may create hurdles for smaller or independent developers, possibly leading to disparity between reported practices by developers with more substantial means and those operating under more constrained circumstances. 

As for the platform, the Checks service could fulfill several objectives for the Play Store. It might elevate the quality of reported practices and compliance of apps. However, Google could face criticism for monetizing a critical aspect of the app development process, particularly if this act is seen as establishing a 'pay-to-play' mechanism.

The implications for users, albeit less direct, are still significant. If the Checks service contributes to higher policy compliance and reduced errors by developers, users could get accurate data safety practices. Conversely, if the cost of the service results in a less diverse app marketplace due to financial barriers for smaller developers, users may face a reduction in their choices of apps. \smallskip

\noindent
\textbf{Regulators}
Even though our analysis finds inconsistencies between privacy labels and privacy practices, evidence ~\cite{zhang2022usable} suggests that privacy labels have can carry specific information about the practices. This information can be very useful for privacy concerned users who want to ensure that they are only using privacy respecting apps.

However, as shown in this work, the accuracy of privacy labels is not guaranteed. While developers are required to disclose their data practices in order to obtain a privacy label, there is no guarantee that the information they provide is accurate or complete. 
Therefore, it is necessary to have systems in place to verify the accuracy of privacy labels and to hold developers accountable for any discrepancies. This is particularly important because the false labels can create a false sense of security among the users. 

One potential model for regulating privacy labels is a system similar to the one used for food nutrition labels, which are regulated by the Food and Drug Administration (FDA). A regulatory body could be established to oversee privacy labels and ensure that they are accurate and consistent. This could help to build trust among users and encourage developers to be more transparent about their data practices.

Another solution could be providing the monitoring system that is used during the app submission reviews directly to the user's device. The system could perform real-time dynamic analysis on the app installed on the user's device and show the analysis results, such as network traffic and the data or sensor access logs. This way, the users are given a more detailed view of the privacy practices of the app and would be able to make a better-informed decision on whether to use the app. An example in production includes the Apple \textit{App Privacy Report} introduced in iOS and iPadOS 15.2, which provides data and sensor access logs from apps installed on the device, and network traffic from the apps directly to the users \cite{apple_app_privacy_report}.

\subsection{Limitations}
\noindent
\textbf{Data Collection}
Our data collection process was carried out over specific periods, and thus may not fully capture the dynamic nature of the app ecosystem. Apps can update their DSS at any time, and changes outside of our data collection windows would not be reflected in our analysis. The apps included in our study are a subset of all available apps on the Google Play Store. Although we made an effort to include popular apps and span various categories, our sample may not be fully representative of the entire app ecosystem. Finally, we noted discrepancies in data practices based on the presence of third-party libraries. However, some instances of third-party library usage might have been missed due to limitations in the tools used for detection, leading to potential underestimation of their prevalence and impact. \smallskip 

\noindent
\textbf{Developer Study}
Our developer study relies on email communication with app developers. While this allowed us to reach a larger number of developers, it may have biased our results toward those who were more willing to respond,  potentially neglecting the perspectives of developers who did not reply. 

The nature of self-reported data could also pose challenges to the reliability and accuracy of the information collected. There is the potential for bias in the developers' responses, as they may present information in a way that portrays their apps more favorably. Another limitation of our study is the lack of demographic data about the developers who responded to our email. As we prioritized respecting privacy and ensuring anonymity, we did not collect any personally identifiable information, including demographic details such as age, gender, nationality, or years of experience in app development. Such demographic data could potentially provide meaningful context and allow for a more nuanced understanding of developers' perspectives. For instance, a developer's experience level or geographical location might influence their understanding of privacy issues or their familiarity with privacy regulations. Despite the limitations outlined, our research sheds light on the challenges that the developers face, and it fills a gap in the literature about the factors affecting developers' making while working with the privacy labels. \smallskip

\section{Conclusion}
In conclusion, our study takes a comprehensive approach to examine the landscape of Data Safety Sections (DSS) in Google Play Store apps. Through a large-scale analysis, we highlighted inconsistencies in reported practices, as well as instances of both underreporting and overreporting. 
Our longitudinal study emphasized the dynamic nature of DSS implementation. We observed the persistence of overreporting trends and revealed how developers are still adjusting to the requirements, even ten months after Google's imposed deadline. We investigate the developers' perspective by performing a qualitative study by communicating with over 3500 developers. Our analysis uncovers the process developers undertake when navigating the DSS landscape. We outlined the myriad challenges developers face, the strategies they employ to comply with DSS policies, and the factors prompting changes in their DSS. Finally, based on the challenges and developer behaviors, we provide recommendations aimed at improving the developer experience with Data Safety Sections.

\newpage

\newpage
\bibliographystyle{IEEEtranS}
\bibliography{references, refs.bib}

\newpage
\section{Appendix}
\subsection{Changes in DSS}
\label{app:cdf}

The above figure shows the frequency with which apps change their DSS between the period of June 20, 2022, and May 31, 2023. While the majority of apps did not have change labels, a significant number, 283K (40\%) changed labels at least 1 time. Moreover, there were over 1.3K apps that changed their DSS at least 3 times.

\subsection{Data Practices in Privacy Labels}
\label{app:figures}
\begin{figure*}[ht]
  \centering
  \includegraphics[width=1\textwidth]{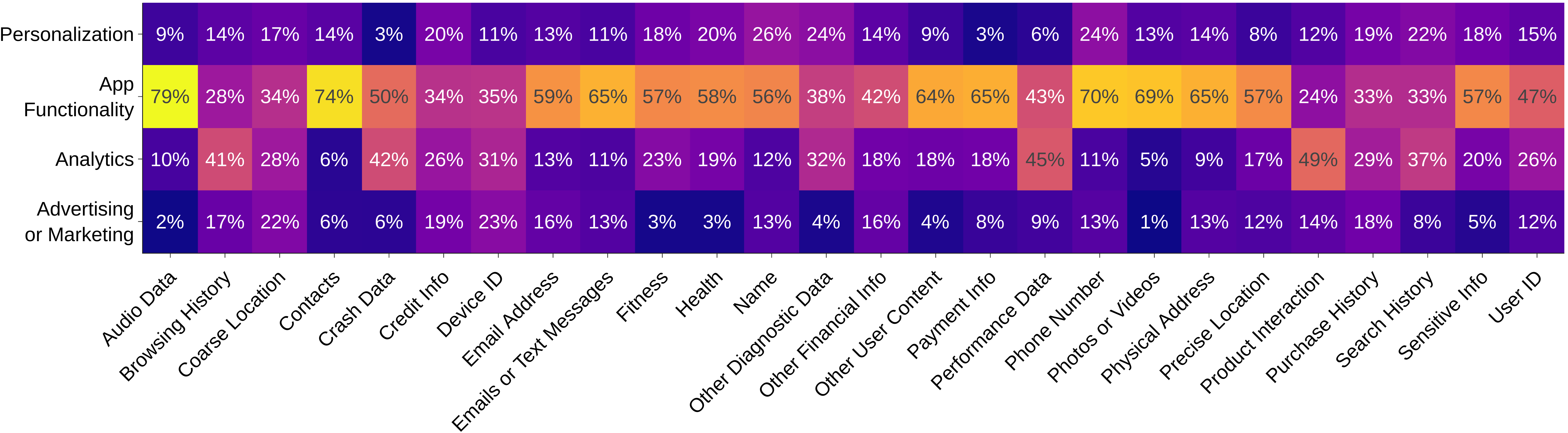}
  \caption{Heat map showing the distribution of data types across purposes.}
  \label{fig:heatmap}
\end{figure*}

In this section, we look into the distribution of the high-level practices of apps based on their price, as shown in \cref{fig:price_high_level_app}. Next, we look at the heatmap in \cref{fig:heatmap} to see the distribution of datatypes across purposes. The heatmap is normalized by the total number of apps collecting or sharing a given datatype. 
\begin{figure}[ht]
  \centering
  \includegraphics[width=\columnwidth]{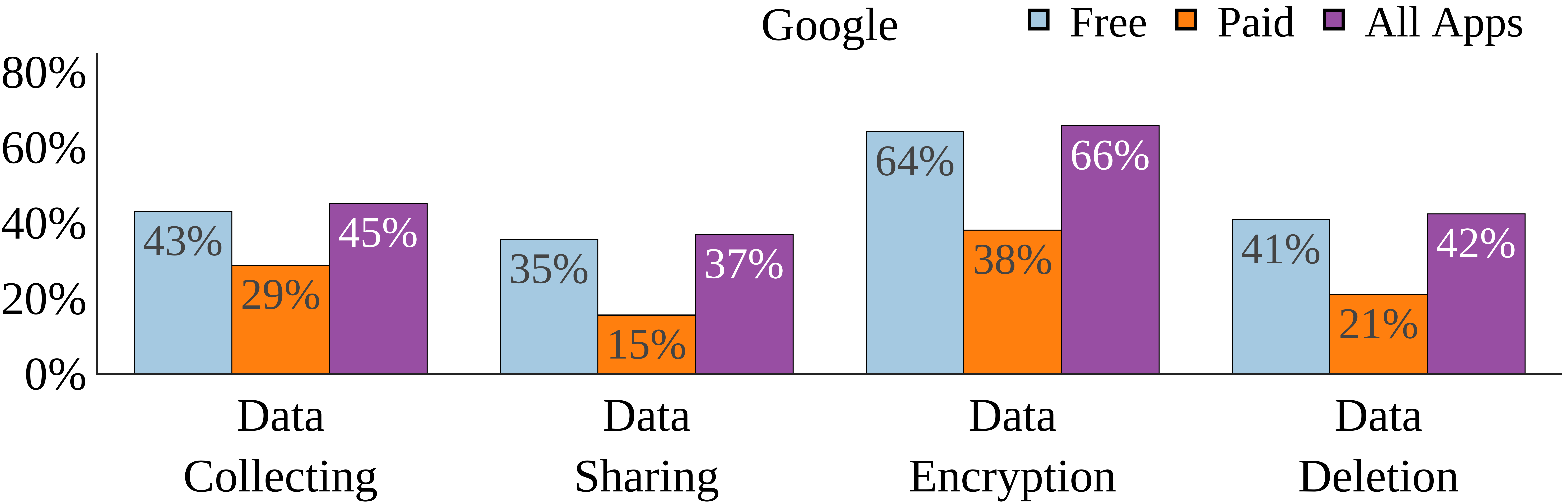}
  \caption{The distribution of high-level data practices based on price.}
  \label{fig:price_high_level_app_app}
\end{figure}

\subsection{Developer Study}
\label{app:dev-study}

For the Developer Study (\cref{sec:developer}) we sent emails to developers in 3 different categories: (A) apps stating that they encrypt data without collecting or sharing data, (B) apps changing their practice from not collecting/sharing data to collecting/sharing data, and (C) apps changing their practices from collecting/sharing data to not collecting/sharing data.

For category (A) we used the following template:
\begin{enumerate}
\item[] \textit{We hope this email finds you well. We are researchers at <LAB\_NAME> and have been using your app, <APP\_NAME>, in our recent studies. We have noticed that in the data safety section of your app, it states that you encrypt data. However, we have also noticed that your app does not collect or share data.}

\textit{We are reaching out to ask if you could clarify this for us. We are trying to better understand the data safety section implemented in your app. We appreciate any information you can provide.}

\textit{Thank you for your time and we look forward to your response.}
\end{enumerate}

\end{document}